# Improving the effectiveness of introductory physics service courses: Bridging to engineering courses

Stephen W. Pierson
*Department of Physics, Worcester Polytechnic Institute (WPI)*
*Worcester, MA 01609-2280*

Suzanne T. Gurland and Valerie Crawford
*Department of Psychology, Clark University, Worcester, MA*

## Abstract

A simple model to help students see the connections between a physics course and their engineering major is described. "Bridging" had positive effects on performance, attitude, and motivation, as measured by grade analyses, surveys, and student and faculty comments. In addition to the improved performance, enrollment in the physics course was also found to increase considerably during the two years of the bridging. The bridging model can be applied to a variety of courses and suggested components are discussed here.

## I. INTRODUCTION

Students often complain that the courses they take have no relevance to one another or to their major, even when the overlap of material can be significant. Perhaps the most overlap of material occurs between physics, math, and engineering courses. Such complaints are understandable given the students' limited perspective and the variety of approaches and notations used to teach the same material.[1] The instructors of such courses can also be part of the problem when we fail to acknowledge the connections.

There have been many attempts to address the situation using an "integrated" curriculum.[2,3,4] In such an approach, a learning module,[5] a project,[6] an entire course,[1] a series of courses, or even back-to-back classes[7] that integrate math, physics and/or engineering are used to emphasize the connections between these fields and "integrate" them. Such approaches typically involve the coordination between faculty from various departments to design the curriculum, schedule the courses, and ensure that the coordination continues through the whole course. While these approaches have met with a range of success, it has also been observed that the integrated approach takes more time and therefore may be deemed as cost-ineffective by the administration.[8]

In this paper, we describe a "bridging" model, a relatively simple approach to help students appreciate connections between courses in different departments. The model, which we applied to a physics course on oscillations and waves and a companion electrical engineering (EE) course, was implemented primarily in the physics course where the EE students were isolated in a single recitation section. There, discussion, projects, homework, and examples emphasized connections to EE material. More limited "bridging" was also done for other majors via projects and, to a lesser extent, homework. The expected outcome for the bridging was that the students would come away with a better and deeper understanding of the physics course and the material from their own major. We also expected that their interest in the physics, and therefore their motivation and attitude, would be improved because of the emphasis on the relevance of the physics material to their major. Further, the understanding of the related material in their major would be enhanced because the material would be presented from a different perspective and in a wider context.

**II.   COURSE INFORMATION AND BRIDGE MODEL**

The bridging centered on WPI's physics course, Oscillations, Waves and Optics (PH1140), which was linked for the EE students to Fundamentals of Electrical and Computer Engineering (EE2014). The bridging, which took place over two consecutive years (1998 and 1999), was expanded to an extent to include all students in the second year.

PH1140 is the fourth course in the introductory physics series. During the *seven-week term*, the course meets five times a week, three of which are conducted in a lecture format and two in recitation sections. In addition, there are three one-hour labs. While the course is taught in both the second and fourth term of the year, the bridging was done in the second term, and thus enrolled almost no freshmen because it is out of sequence. The approximate make-up of the students during the two years of this project was 45% EE, 30% mechanical engineering (ME), 10% computer science, and 10% physics majors, with the rest being made up of chemistry, civil engineering, biology, and math majors.

The first half of the course is spent on oscillations, covering the topics of free, damped, and forced oscillations. The next sixth of the course covers the basic terminology, types, and properties of waves, while the final third of the course applies the previous material to sound and optics. Because oscillations, along with the complementary idea of resonance, are ubiquitous in nature and technology, the course is ideal for bridging. For example, the ME majors can apply this information to machines while the EE majors can apply it to circuits. The EE majors can also apply the material on waves to power transmission in systems ranging from small devices to power lines.

EE2014 is an EE majors course typically taken by sophomores. The course is a natural bridging partner for PH1140 because it is an introduction to differential equations with applications to $1^{st}$ and $2^{nd}$ order systems, including resistor-capacitor-inductor (RCL) circuits and driven RCL circuits. Phasors and transients are also treated in both courses. Mirroring the second half of PH1140, EE2014 covers signal propagation in various systems, including transmission lines, along with other material that was not bridged. In both years of the bridging, the author SWP taught PH1140 while different instructors taught EE2014.

In the first year 1998, the bridging was for the oscillation portion of the course (i.e., roughly the first half) and only for the EE majors, who were separated into a separate recitation section and most of whom were also taking EE2014. The bridging model for the EE majors had several components. The most significant change was to put the EE majors into a separate recitation section, known as the Bridge section, in which an emphasis was placed on EE applications in the examples and discussion. We also strove to explain the differences in approach and notation of the two disciplines for the overlapping material. For the weekly homework assignments, one of the five problems was specific to circuits and a small discussion of the application of the material to their major was included. Further, for this section, an EE professor was brought in to guest lecture on the relevance of the material to their major. Finally, all of these students were required to complete a project on resonance in resistor-capacitor-inductor circuits using phasor analysis. The project was a five-page handout laying out the problem, reiterating some theory, and ending in a series of questions. The project was estimated to be the time-equivalent of one and a half of the weekly homework assignments.

In the second year 1999, the bridging was expanded to all students in the course in two ways. First, as part of the initial homework assignment, each student was required to ask a professor from the major department the significance of the physics course to their major. Second, each student was required to do a project in one of three areas, EE, ME, or biology, depending on their major. Students in the course who were not one of these majors were free to choose any of the three projects. All three of the projects were designed to be general enough to be useful to a student of any major. (Sample projects available at http://www.WPI.EDU/Academics/Depts/Physics/News/Talks/Pierson/bridge.html.) The ME project focused on resonance and oscillations in an unbalanced rotor, which has applications on topics ranging from car's tires to washing machines to propellers and turbine blades. The biology project centered on the physics of the ear, which can be thought of as a forced oscillator. The EE project was an elaborated version of that from the first year. The bridge for the Bridge section was also expanded in the second year to include the wave portion of the course, in addition to the oscillations portion covered the first year. In both years, recruiting for the EE Bridge Section was done through flyers, announcements in the EE majors course that precedes EE2014, and email messages to students registered for EE2014, EE students already registered for PH1140, and EE advisors.

**III. EVALUATION**

As stated above, our hypothesis was that the bridging would produce a variety of performance-oriented, motivational, and attitudinal effects on students exposed to the bridge treatment. To evaluate the success of the Bridge, the Bridge group and appropriate Non-bridge control groups in the two courses were compared on exam and attitude measures. Here, we will outline the evaluation procedures, referring the reader to the full reports[9,10] for the details.

The Bridge group is defined as the group of EE students cross-enrolled in the PH1140 EE section and EE2014. Note that the PH1140 EE section was not made up exclusively of Bridge students but included other EE majors as well as non EE-majors that could not work the other section into their schedule. Our evaluation focused on the Bridge group, despite the fact that the bridging was partially expanded to all students in the second year.

To evaluate the Bridge group's learning of the content, two independent control groups were formed, one for each course. The Bridge students' scores on exams or quizzes covering bridge-relevant material were then compared to that of the comparison group of Non-bridge students for each course. In the first year, the comparison groups were assembled to the extent possible by matching to each Bridge student a randomly selected Non-bridged student from each course, so that self-reported SAT Math scores of the two students were within 10 points of each other, and academic year was the same or one year different. In the second year, the matching was based on both a high school GPA within 0.2 of the Bridge student's high school GPA, and a college GPA within 0.2 of the Bridge student's college GPA. Statistical tests were also performed that year to ensure the two groups in each course were statistically equivalent in baseline school performance with virtually identical GPAs. These measures were taken so that any differences in performance between the Bridge and Non-bridge groups could be rightfully attributed to the bridging intervention and not simply to chance differences in baseline school performance.

The two groups were also compared on attitude. In both years, attitudes toward cross-disciplinary work and their enjoyment and overall rating of the course were evaluated using pre-course and post-course questionnaires that were designed to measure the effects of bridging treatment on students' attitudes and expectations about conceptual linkages between disciplines. In the second year, recitation section attendance and rate of homework completion were compared for the two groups to assess motivational issues.

The primary statistic used in our analysis of the results is the F-statistic.[11] Like the t-statistic, it is used to compare group means in order to judge whether a treatment had an effect. It is derived by analyzing the variability (variance) of scores within each group and between the groups. The result is a partitioning of the variances of all scores within and between groups into variance due to "error" (i.e., variance that is unexplained or unattributable to any variable being explored) and variance due to the "effect" (i.e., variance attributable to the effect of treatment). This process of comparing differences between means by examining variances is called "analysis of variance" (ANOVA). Thus the F statistic represents a ratio of explained or systematic variance to unexplained (i.e., random) variance.

| EE2014 | Year | N | Exam Total | Bridge | Non-bridge | Score difference | F value | p-value |
|---|---|---|---|---|---|---|---|---|
| Quiz 2* | 1998 | 7 | 100 | 57.1 | 50.7 | +6.4 | 0.33 | 0.57 |
| Exam 2* | 1998 | 7 | 100 | 69.6 | 55.3 | +14.3 | 0.99 | 0.34 |
| Exam 1 | 1999 | 14 | 100 | 70.8 | 71.1 | -0.3 | 0.00 | 0.95 |
| Exam 2* | 1999 | 14 | 100 | 86.6 | 80.5 | +6.1 | 1.75 | 0.20 |
| Exam 3 | 1999 | 14 | 100 | 74.8 | 70.0 | +4.8 | 0.51 | 0.48 |
| Exam 4 | 1999 | 14 | 100 | 83.7 | 85.5 | -1.8 | 0.17 | 0.69 |

**Table 1:** Means and ANOVA results for Bridge and Non-bridge groups' exam scores in EE2014 for both years. The starred (*) measures are the ones that covered bridge materials.

F-scores are evaluated (tested) by checking the probability ("p" or "p-level") that the observed F-score would occur "under the null hypothesis", i.e., if one assumes that the treatment had no systematic effect. Conventionally, behavioral scientists agree that a probability of 0.05 (also called the "alpha level" or "criterion") is sufficiently small to conclude that the observed ratio (the F-score) is not simply a chance occurrence.

## IV. RESULTS

During the two years of the bridge, the number of "Bridge Students" increased from 8 (out of 53) to 15 (out of 67). The number of students in the PH1140 EE section also increased from 10 to 28. While the bridge was assessed for both motivation and performance, a number of other results will be discussed as well.

**4.1 Performance and motivation**

Our analysis points towards a systematic improvement of both performance and motivation. Although the number of students was too small to reach statistical significance, we believe that the improvements would be detectable with increased sample size or with more sensitive measurement techniques. The data show a consistent pattern of Bridge students performing better than the control group on the exams that cover bridged material. We will present the main results here, referring the reader to the full reports[9,10] for the details.

As can be seen in Table 1, the scores for the EE2014 exams or quizzes that covered bridged material were consistently higher for the Bridge group. The Bridge group scored between 7% and 25% better than the Non-bridge students did. As a comparison, the scores for the other three exams in the 1999 EE2014 course where bridging material was only a minor part are also included in that Table.[*] As one can see, the scores are much closer there, indicating that any gained performance is limited to the bridged material. Yet, this finding also discounts our expectation that the improved performance of the Bridge students would carry over to non-bridged material due to increased motivation and interest.

Similar but less pronounced results were found for most of the PH1140 exams, as shown in Table 2. The exception is the first exam from 1998 where the Bridge group scored marginally less than the comparison group. For the 3$^{rd}$ exam in 1998 and 1999 where little bridged material

| PH1140  | Year | N   | Exam total | Bridge | Non-bridge | Score Difference | F value | p-value |
|---------|------|-----|------------|--------|------------|------------------|---------|---------|
| Exam 1* | 1998 | 8   | 62         | 44.6   | 47.8       | -3.2             | 0.43    | 0.52    |
| Exam 2* | 1998 | 8   | 68         | 50.8   | 48.3       | +1.7             | 0.28    | 0.60    |
| Exam 3  | 1998 | 7+  | 65         | 54.7   | 55.4       | -0.7             | 0.02    | 0.88    |
| Exam 1* | 1999 | 14  | 75         | 57.2   | 48.1       | +9.1             | 4.00    | 0.06    |
| Exam 2* | 1999 | 14  | 75         | 57.9   | 54.6       | +3.3             | 0.85    | 0.37    |
| Exam 3  | 1999 | 14  | 75         | 65.1   | 63.9       | +2.2             | 0.17    | 0.68    |

+One of the Bridge group students did not complete the third exam and so the matched companion from the control group was omitted from the analysis of the Exam 3 score.

**Table 2:** Means and ANOVA results for Bridge and Non-bridge groups' exam scores in PH1140 for both years. The starred (*) measures are the ones that covered bridge materials.

was covered, the differences were somewhat less pronounced, lending some credibility to the evidence that the bridging improved the performance of the Bridge students in PH1140.

While we believe that the improved performance of the Bridge students over the control groups is due to the bridging, there may be other factors to help explain the difference in scores for the Bridge group and the comparison group. The most likely factor is the student to teacher ratio for the recitation sections, at least for the first year. In PH1140, that ratio was 10:1 for the Bridge section and 43:1 for the non-Bridge section the first year and 28:1 and 39:1, respectively, the second year. It is unlikely that the student to teacher ratio was as influential in the second year. (Note that EE2014 did not meet in sections and so the student to teacher ratio was not a factor there.)

One could also argue that the Bridge students performed better on their EE2014 exams and quizzes because of the repetition of material between the courses. Yet this is also part of the purpose of the bridging: for students to see the similarities in material for two courses. Another possible explanation for the improvements seen by the Bridge students is that they are a self-selected group. In recruiting for participation in the bridging, it may be that students who chose to enroll were better in ways not reflected in our selection criteria for the comparison group.

---

* The equivalent analysis was not done in 1998 and the data are no longer available.

| Index | Bridge | Non-bridge | F value | p-value |
|---|---|---|---|---|
| Conference Attend. | 0.86 | 0.74 | 2.69 | 0.11 |
| Missed Homework | 1.93 | 0.29 | 6.01 | 0.02 |

**Table 3:** Means and ANOVA results for Bridge and Non-bridge groups' motivational indices in PH1140.

Offsetting the student/teacher ratio for the Bridge students in PH1140 was the exam material. To minimize any advantage of the Bridge students on the PH1140 exams, no EE-related problems were included on the exams. (I.e., only problems regarding oscillations or resonance in mechanical systems and not RCL circuits appeared on the exams.) Further, because the EE students had more homework problems related to EE material, it could be argued that they had a slight disadvantage on the exams because they were doing fewer homework problems related to the exam problems. It is also possible that the bridge projects for all students in the second year minimized the differences between the Bridge group and Non-bridge group in PH1140. Unfortunately, such differences could not be untangled in our assessments.

Two indices were used to compare the Bridge and Non-bridge students on motivational variables in the second year of the bridging. Specifically, the groups were compared, using an ANOVA framework, on their conference attendance and on their number of missed homework assignments. The hypotheses were that Bridge students would have greater learning motivation, and that this motivation would be evidenced by greater conference attendance and fewer missed homework assignments. As indicated in the Table 3, the difference in conference attendance did not reach statistical significance. It was, however, in the predicted direction, with Bridge students attending an average of 86% of the conferences, and Non-bridge students attending an average of only 74%.

The findings for missed homework assignments are rather enigmatic. The difference between the two groups did indeed reach statistical significance, as indicated in Table 3, however it was in a direction opposite that which was hypothesized. Specifically, while Bridge students missed an average of 1.93 homework assignments, Non-bridge students missed an average of only 0.29. One possible, though counter-theoretical, explanation for this finding is that the bridge treatment actually lowers student motivation. An alternative explanation arises, however, when this finding is viewed in light of the Bridge students' higher exam scores. Given that the Bridge group consistently out-scored the Non-bridge group on exams, despite having completed fewer of the homework assignments, we might suggest: 1) that the Bridge students felt more confident in their conceptual understanding of the material and therefore did not need to make as much use of homework assignments to further their learning; and/or 2) that the Bridge students would have out-scored the Non-bridge students by even more points had they completed a greater proportion of the assigned homework.

In both years of the bridging, a questionnaire was used at the beginning and end of PH1140 to evaluate whether bridging of course content affected students' attitudes regarding the inter-relatedness of introductory courses across disciplines, enjoyment of the course, and overall rating of the course. (The questionnaires are available in the full reports.[9,10]) In the first year, while the Bridge group did score higher than the Non-bridge group on the post-course questionnaire as predicted, the difference was not statistically significant ($F = 1.73$, $p = 0.21$). (See Table 4.) But more important, both groups' questionnaire means fell from pre- to post-survey score. The Bridge group pre-course questionnaire mean was higher than the comparison group mean, but it fell about 1.8 points less than did the comparison group's score.

| PH1140 | Year | N | Pre-course Survey means | Post-course survey means |
|---|---|---|---|---|
| Bridge | 1998 | 6 | 17.00 | 16.33 |
| Non-bridge | 1998 | 8 | 14.75 | 12.25 |

Table 4: Means for Bridge and Non-bridge Groups in PH1140 on Attitudes toward Interdisciplinarity Questionnaire. The range of scores possible is [-30:30].

The questionnaire results can be interpreted in two ways. It is possible that the Bridge group students scored higher on the pre-course questionnaire because, in part, they were told that they were part of a project that would be promoting links between two courses, and this positively influenced their expectations and attitudes toward interconnections among disciplines. Further, it is possible that the bridging experience did not match their expectations, and therefore their questionnaire scores fell rather than rose. Nonetheless, our results indicate that that the bridging may buffer students from attitude slippage.

In addition, it is possible that in general, students at the beginning of a course have a tendency to complete surveys related to their expectations with more general goodwill and optimism than at the end of a course. That both groups rated their satisfaction with the course virtually identically (data not shown), as measured by the post-course questionnaire suggests that this general lessening of optimism is not related to students' experiences with the bridging treatment per se or with satisfaction with the course, which was quite high (data not shown).

In the second year, the Bridge and Non-bridge groups' reported attitudes, course enjoyment, and overall course rating at the end of the term were compared (data not shown). (An administrative oversight prevented the pre/post comparison from being made.) Contrary to the first year, it was determined that the Bridge and Non-bridge groups did not differ significantly in their attitudes toward cross-discipline inter-relatedness. Similarly, the two groups did not differ with regard to course enjoyment, nor with regard to overall course rating. We attribute much of the minimal difference in attitudes in the second year to the fact that all students in PH1140 had to do bridging projects.

**4.2 Aspects not assessed**

During the course of the bridging, we noticed a variety of benefits beyond those assessed, ranging from class enrollment to interdepartmental communication. Even though these observations were not quantified, we believe that their value is real and should be noted here.

From the year before bridging started to the second year of bridging, PH1140 enrollment increased from 39 to 67, an increase of nearly 70%. While many factors could explain this, word of mouth about the bridge as well as the recruiting for the bridge are believed to be the primary factors. In a time of declining physics enrollments, such increases in enrollment are important.

An examination of student comments on the standard WPI course evaluation forms reinforced the value of the bridging. While there were no negative comments about the bridging, there were over a dozen positive comments. (Many of the comments appear at http://www.WPI.EDU/Academics/Depts/Physics/News/Talks/Pierson/bridge.html.) The students described the bridging as not only "useful", helpful, and "well implemented", but also "enjoyable" and "very enlightening." Other students described how much they liked the bridging, what an "excellent" or "good" idea it was, and how "great" it was. The common theme seemed to be that the students valued seeing the connection of the physics course to their major or a particular class that they were taking.

Other significant benefits of bridging included the exposure of the course instructor to the courses taught in other departments and the interests of those departments. As pointed out in Reference [1], knowing how the physics is used in other courses can significantly aid in helping the students to understand and apply what is being taught in the physics course. Another benefit that seemed vital in the bridging was that differences in notation between disciplines were explained to the students so that such differences would not be a stumbling block for them. With physics departments at engineering schools being largely service departments, it is important for them to maintain contact with the departments whose students they serve. Such interdepartmental communication is especially important to maintain physics course populations as more engineering departments teach their students physics the way they would like it taught and require fewer courses from a physics department.

## V.  DISCUSSION AND SUMMARY

Overall, the analysis of our bridging indicates that the bridging had a systematic influence on performance and attitudes for the students. Although not to a statistically significant degree, the Bridge students tended to score better on exams or quizzes where bridge material was covered. The students also seemed to better appreciate the conceptual linkages between the physics and EE disciplines. Further, the strong support for the bridging from the student course evaluations reinforced the value of the bridging. Anecdotally, subsequent comments to the instructor of PH1140 (SWP) months or years later confirmed that the importance of the physics course to the students' discipline was conveyed to the students.

In evaluating this bridge model and others at WPI, a group of WPI faculty identified what we believed to be the principal components of a successful bridge.[12] Those components were: (i) "contextual learning", i.e., putting the concepts that the student is taking in one course into the terms of another course; (ii) an outside expert, typically a guest lecturer from outside the department offering the course, who gives a presentation for part or all of a class; (iii) pinpointing possible confusions in style, terminology, and notation for the student; and (iv) communication between the instructors of the two courses. While the successful components were speculative, they were largely backed up by anecdotal comments of the students. The first component seemed to be important because it gave the students an identity. By putting the student into a group or section according to their major, the instructor was acknowledging their major and emphasizing to the student that the material they were learning in that course is important to their major. The second component, an outside lecturer, tended to give credibility and authority to the importance of the bridging. It also reinforced the connections between that course and the discipline of the outside lecturer. In terms of the third component, most students could recognize that the concepts in two different courses were similar but, without bridging, were unable to take advantage of the connections because of the differences in approach. The fourth component not only helped the instructors identify various areas of confusion for the students, but it put into practice what the instructor was teaching: communication and connections between disciplines. This component had the most impact when students see or know of the communication between the instructors.

**ACKNOWLEDGEMENTS**

The authors gratefully acknowledge the support of the NSF Institution Wide Reform Project at WPI (NSF # DUE-9653707). The input and support of Judy Miller is especially appreciated. SWP would also like to acknowledge the enthusiastic cooperation from the following members of the EE department: Imad Nejdawi, Willie Eggiman, John Orr, and Len Polizzotto.

---

[1] J. W. Dunn and J. Barbanel, *"One model for an integrated math/physics course focusing on electricity and magnetism and related calculus topics,"* American Journal of Physics, **68** (8), 749-757 (2000).

[2] A. McKenna, F. McMartin, A. Agogino, *"What students say about learning physics, math, and engineering,"* 30th Annual Frontiers in Education Conference, vol.**1** (IEEE Cat. No.00CH37135) p. T1F/9, 2000.

[3] J.E. Froyd, G.J. Rogers, *"Evolution and evaluation of an integrated, first-year curriculum,"* 27th Annual Frontiers in Education Conference, vol.**2** (IEEE Cat. No.97CH36099) p. 1107-1113, 1997.

[4] J.R. Morgan and R.W. Bolton, *" An integrated first-year engineering curricula,"* 28th Annual Frontiers in Education Conference, vol.**2** (IEEE Cat. No.98CH36214) p. 561-565, 1998.

[5] S. Anwar, *"Development of integrated learning modules for engineering technology students,"* 30th Annual Frontiers in Education Conference, vol.**2** (IEEE Cat. No. 00CH37135) p. S2E/5, 2000.

[6] R.J. Roedel, D. Evans, R.B. Doak, J. McCarter, S. Duerden, M. Green, and J.Garland, *"Projects that integrate engineering, physics, calculus, and English in the Arizona State University Foundation Coalition freshman program,"* 27th Annual Frontiers in Education Conference, vol. **1**, (IEEE Cat. No. 97CH36099), p. 38-42, 1997.

[7] E.W. Hansen, *"Integrated mathematics and physical science (IMPS): a new approach for first year students at Dartmouth college,"* 28th Annual Frontiers in Education Conference, vol. **2**, (IEEE Cat. No.98CH36214), p. 579, 1998.

[8] T. Shumpert and P. Zenor, *"Auburn University integrated pre-engineering curriculum (IPEC): a half-time report,"* 28th Annual Frontiers in Education Conference, vol. 2, (IEEE Cat. No.98CH36214), p. 830, 1998.

[9] V. M. Crawford, *Bridge Project: PH1140, Oscillations, Waves & Optics and EE2014, Fundamentals of Electrical and Computer Engineering,* WPI IWR Project Evaluation Report, 1999 [Available at http://www.WPI.EDU/Academics/Depts/Physics/News/Talks/Pierson/bridge.html]